\documentclass[conference]{IEEEtran}
\pdfoutput=1
\IEEEoverridecommandlockouts
\usepackage{cite}
\usepackage{amsmath,amssymb,amsfonts}
\usepackage{algorithmic}
\usepackage{graphicx}
\usepackage{textcomp}
\usepackage{todonotes}
\usepackage{xcolor}
\def\BibTeX{{\rm B\kern-.05em{\sc i\kern-.025em b}\kern-.08em
    T\kern-.1667em\lower.7ex\hbox{E}\kern-.125emX}}
    
\usepackage{etoolbox}
\usepackage{xstring}
\DeclareListParser{\doslashlist}{/}
\newcounter{ndnNameComponentCounter}%
\newcommand{\name}[1]{{%
  \setcounter{ndnNameComponentCounter}{0}%
  \renewcommand{\do}[1]{{%
    \ifnumgreater{\value{ndnNameComponentCounter}}{0}{\allowbreak/}{}%
    \ifnumodd{\value{ndnNameComponentCounter}}{}{}%
    ##1}%
    \stepcounter{ndnNameComponentCounter}}%
``{\fontfamily{cmtt}\small\selectfont\IfBeginWith{#1}{/}{/}{}\doslashlist{#1}}''%
}}    
    
\usepackage[absolute]{textpos}

\begin{document}

\begin{textblock*}{\textwidth}(1cm,1cm)
\noindent\large \colorbox{red}{This is the accepted version of the paper. The final version will appear in the proceedings of ACM SACMAT 2022.}\\
\end{textblock*}

\title{AABAC - Automated Attribute Based Access Control for Genomics Data}

\author{\IEEEauthorblockN{David Reddick}
\IEEEauthorblockA{\textit{Computer Science} \\
\textit{Tennessee Tech University}\\
Cookeville, TN, USA \\
dereddick42@tntech.edu}
\and
\IEEEauthorblockN{Justin Presley}
\IEEEauthorblockA{\textit{Computer Science} \\
\textit{Tennessee Tech University}\\
Cookeville, TN, USA \\
jcpresley42@tntech.edu}
\and
\IEEEauthorblockN{F. Alex Feltus}
\IEEEauthorblockA{\textit{Department of Genetics \& Biochemistry} \\
\textit{Clemson University}\\
Clemson, SC, USA \\
ffeltus@clemson.edu}
\and
\IEEEauthorblockN{Susmit Shannigrahi}
\IEEEauthorblockA{\textit{Computer Science} \\
\textit{Tennessee Tech University}\\
Cookeville, TN, USA \\
sshannigrahi@tntech.edu}
}

\maketitle

\begin{abstract}
The COVID-19 crisis has demonstrated the potential of cutting-edge genomics research. However, privacy of these sensitive pieces of information is an area of significant concern for genomics researchers. The current security models makes it difficult to create flexible and automated data sharing frameworks. These models also increases the complexity of adding or revoking access without contacting the data publisher. In this work, we investigate an automated attribute-based access control (AABAC) model for genomics data over Named Data Networking (NDN). AABAC secures the data itself rather than the storage location or transmission channel, provides automated data invalidation, and automates key retrieval and data validation while maintaining the ability to control access. We show that AABAC when combined with NDN provide a secure and flexible combination for work with genomics research.
\end{abstract}


\section{Introduction}

As exemplified by the COVID-19 crisis and the subsequent rapid vaccine development, genomics research has the potential to revolutionize healthcare. With computing becoming cheaper and genome sequencing machines becoming ubiquitous, the genomics community is generating a massive number of valuable, geographically distributed datasets. Researchers often desire to share those datasets with the research community and healthcare providers. However, as genomics data begins to become larger and more distributed, an acute problem arises - the complexity of sharing data with other researchers while providing fine grained and easy access control.

Consider this example; a Principle Investigator (PI) wants to share access to a restricted dataset with a new graduate student. With a traditional Public Key Encryption model (PKI), either the data needs to be stored unencrypted in a ``secure" location that only the PI and the students can access or every person needs to have a copy of the data encrypted with their public keys, resulting in multiple copies of the same data. When a new student joins the group, the data must be re-encrypted with their public key, creating yet another copy of the data. Genomics data is rapidly approaching Exabytes, and this approach of creating multiple copies of the data is not sustainable\cite{Qin2020}\cite{Ogle2021}. While traditional attribute based access control methods have been proposed, they suffer from performance bottlenecks and from the complexity of key discovery and retrieval\cite{Abinaya2021Aug}. 
In the genomics community, access revocation is generally archived by revoking access to the storage location. However, access control based on files do not work when superusers have access to all the directories on a system. As the genomics community moves towards the cloud computing model where the hosts and computing platforms are potentially untrusted, the data itself must be secured both in transit \textit{and} at rest. While a large-scale confidentiality breach for genomics has not been documented, it is an active concern for individuals in the field \cite{Phillips2020}. Finally, the act of access control by centralized reencryption and key revocation may not scale.


This work proposes a novel scheme that addresses these problems through an attribute-based access control model supported by Named Data Networking (NDN). We have worked with domain scientists to better understand their requirements.
Our contributions are threefold (a) we propose an attribute based encryption scheme that invalidates data after a certain time, enabling time-based control access (b) we propose a hybrid access model using the combination of local and remote ledgers that allow both data publisher as well as institutional access control over published data, a key requirement for the genomics community, and (c) we automate and simplify key discovery, delivery, and verification based on the content names. We utilize the name based primitives of NDN that contain content names for all operations. Our approach has several advantages over traditional methods. First, we allow both the publisher and trusted collaborators (such as an administrator) to control access to data. However, unlike today, the administrators do not gain access to the data. 
For example, when a student graduates and no longer needs access to the data, the university can revoke access without involving the publisher. Second, in NDN, the decryption keys are linked to the data itself, automating key retrieval and data decryption. Finally, a time based partial reencryption model maintains confidentiality without incurring a large overhead.

\section{Related Work}
\subsection{NDN Background}

NDN is a future Internet architecture, which bases its communication protocols on content names for data requests rather than traditional networking based on IP addresses of hosts for delivery \cite{Zhang2010} \cite{Zhang2014}. NDN also facilitates other in-network features such as caching, multicast, and location agnostic data retrieval. 
All data is signed at publication and data integrity can be ensured regardless of where data is stored. A human-readable, hierarchical, and application-defined naming scheme is used for the identification of content and in-network operations. This flexibility in naming allows integration with existing naming schemes, such as those used for scientific data like genomic databases \cite{Ogle2021}.  Communication within NDN uses Interest and Data packets. To get Data, the consumer sends an Interest packet that is forwarded based on the content name prefix to a Data source. Once the Interest reaches a data source, the Data packet, segmented into 8800 byte (configurable) pieces, follows the return path. For brevity, we do not discuss NDN in more detail but refer the reader to previous work\cite{Zhang2014}\cite{Ogle2021}.


\subsection{Access Control for Genomics Data}

There have been previous efforts to address access control for genomics data. Brewstar et al. \cite{10.1162/dint_a_00029} has presented an ontology-based access control for distributed scientific data. This work provides access control to data using a role-oriented database that stores data attributes and their relationships. Mukherjee et al.\cite{10.1145/3041048.3041055} talks about Fast Healthcare Interoperability Resources (FHIR), a standard for swift and efficient storage/retrieval of health data. This effort 
uses an Attribute-Based Access Control (ABAC) system to act as a middle layer between the client application and FHIR server to facilitate fine-grained access to FHIR resources.
Narouei et al. \cite{10.1145/3078861.3078874} introduced a top-down engineering framework for ABAC that autonomously retrieved policies from unrestricted natural language documents and then used deep neural networks to extract policy-related data. 
Namazi et al. \cite{10.1145/3297280.3297428} presented work on developing an attribute-based privacy-preserving susceptibility method that outsourced genomics data to an unreliable platform. The computations' challenges were determined to process the outsourced data and grant access concurrently within patient-doctor interactions.

However, none of these works address a crucial gap in access control for genomics data - modern collaborative research. Unlike healthcare settings where data is small in size and potentially shared with a small number of people (e.g., doctors of a patient), research collaborations require flexibility where participants often change and a hybrid access control is desired. 
\section{The Current Data Security Model}
\begin{figure} [!ht]
    \centering
    \includegraphics[trim={1cm 1cm 1cm 1cm},width=0.70\columnwidth]{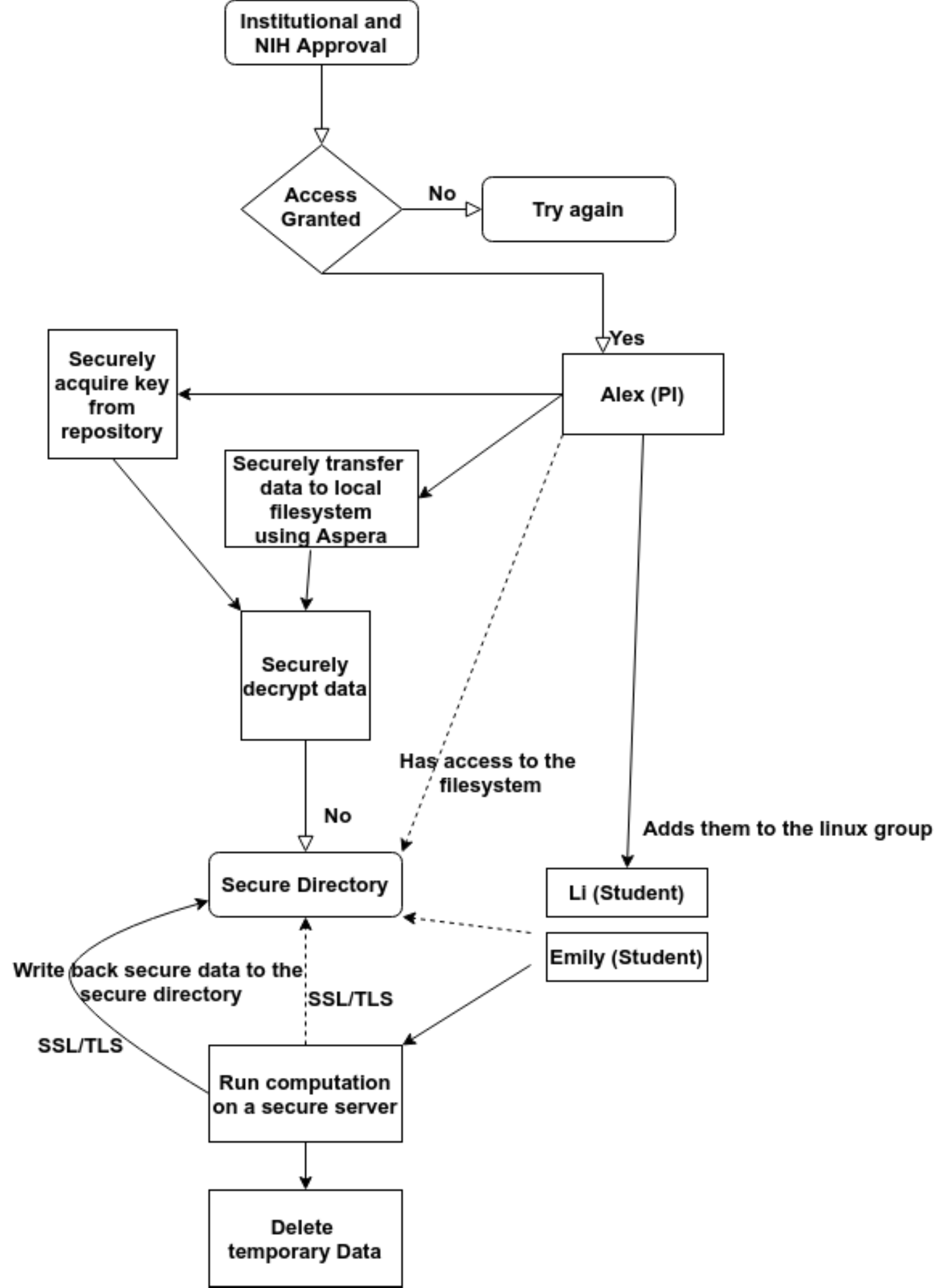}
    \caption{\textbf{Representative Data Security Model for Genomics Data from our collaborators.} The exact method slightly varies based on the institute and research group.}
    \label{fig:current_approval_process}
    \vspace{-1em}
\end{figure}

Figure \ref{fig:current_approval_process} shows a representative data security model currently used by the genomics community. This model was developed in collaboration with genomics researchers. 
While some details might differ based on the actual use case across (sub)communities, the figure should provide a general overview. First, an institutional review board reviews the request for privacy-sensitive data. Once approved, the Principal Investigator (PI) then requests access to the data repository data. Depending on the type of data, this can be hosted at another institute or an institute such as the National Institutes of Health (NIH), National Center for Biotechnology Information (NCBI), Sage Bionetworks, or Broad Institute. The PI needs to name the graduate students and anyone he wants to give access to at the time of this request and add them to the IRB. Once the request is approved, the PI securely transfers data into a secure local location. The location of the data can then be secured in various ways, such as file system permissions, Linux group restrictions, or some custom access control method. If a student needs access to the data, the PI adds the student to the Linux group. When computations need to run on the data, the data is securely transferred (TLS/SSL) over to a secure computational facility. The results are then securely written back to the secure directory. 

Adding a layer of encryption at rest for the data would assist security but is not easily implemented with the current model. One approach is to share a private key among the students and their collaborators, which is not recommended. 
This also complicates the ability to revoke access when a student leaves. The alternative approach is to create a per-person copy of the data - which does not scale.

\section{AABAC Design}

\begin{figure}
    \centering
     \includegraphics[width=0.4\textwidth]{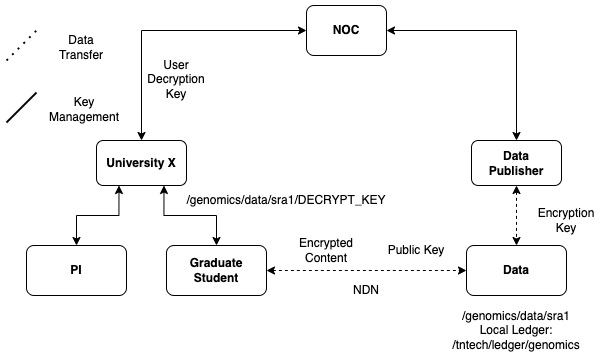}
    \caption{Overview of the system}
     \label{fig:encryption_decyption_overview}
     \vspace{-1em}
\end{figure}


\begin{table*}[ht!]
\caption{NDN names used to facilitate AABAC.}
\begin{center}

    \begin{tabular}{|p{4.8cm}|p{9.7cm}|}
    \hline
    Content Name & /genomics/data/SRA/9605/… \\
    \hline
    Encryption key name & /genomics/pub\_key/sequence=random\_number\\
    \hline
    Local ledger locator & /genomics/data/sra1/$<$attributes$>$/Alice-pub-key:/tntech/alice/pub\_key\\
    \hline
    Name fowarded to NOC & /tntech/ledger/decryption-key/data:/genomics/data/sra1/$<$attributes$>$/Alice-pub-key: /tntech/alice/pub\_key\\
    \hline
    Name of NOC's reply with decryption key & /tntech/ledger/decryption-key\_{alice}/genomics/data/sra1\\
    \hline
    \end{tabular}
    \vspace{-2em}
    \label{tab:nameTable}
\end{center}
\end{table*}

Figure \ref{fig:encryption_decyption_overview} outlines how automated attribute based access control (AABAC) works. 
Once the IRB is approved, the PI's institute and the data publisher agrees on the attributes for data encryption. In this example, the published data is named as "/Genome1/SRA/9605/9609/RNA-Seq/1" and encrypted with mutually agreed-upon attributes such as "PI and PI's graduate students". The data publisher trusts the PI (trust is established through the IRB process) to control access to the sensitive data and does not wish to be involved in issuing or revoking decryption keys. The PI's utilizes his/her university's authentication system to issue or revoke keys. We also assume there is an Network Operation Center (NOC) that both the data publisher and the university trusts. This NOC is in charge of issuing keys to the users as well as publishing the keys for data encryption. Finally, to assist with understanding the NDN naming schemes discussed in this section, Table \ref{tab:nameTable} is provided to outline the names in a dedicated format.


\subsection{Key Generation and Publication}
In AABAC, the NOC is in charge of creating and distributing the master key( $m_k$) and the public key ($p_k$). $p_k$ is used for encrypting data while the $m_k$ is 
used for creating decryption keys. Once these keys are generated, the public key is published into an NDN network where anyone can request and utilize these keys for encryption. Since NDN is location agnostic, the key can be published to a repository or cached in the network. The master key is not published. The public key can be named as \name{/genomics/pub\_key/sequence=random\_number}, where the seq is a random number used to distinguish between different $p_ks$. 

\subsection{Namespace and attribute mapping}
As part of the IRB review, the PI and the data publisher agrees on a set of attributes. In our example, "/genomics/data/" may have attributes "PI=Tom or PI's graduate students". These attributes can also be published into an NDN network under the names "/genomics/data/attributes". The publisher will then be able to retrieve these attributes from the network when encrypting. 

\subsection{Data Publication}
When a file named "/genomics/data/sra" is published, the publisher will request a $p_k$ from the NDN network. This key may be a specific key or a random key provided by the network. The data publication process after this step is simple, the $p_k$ is applied to the data to create encrypted content $e_c$. This encrypted data is also published into the NDN network under a name such as \name{/Genome1/SRA/9605/9609/RNA-Seq/1/encrypted\_by=/genomics/pub\_key/timestamp=1645780366}. Once data is encrypted, they can be  published in an NDN repo that makes the data available for anyone asking for the content. As we discussed earlier, a file in NDN is divided into several Data packets and signed. In AABAC, each data packet is individually encrypted before being digitally signed by the publisher. The signature provides provenance and enables us to publish this data from anywhere. Note that while data is available from anywhere, they are encrypted and can not be used unless the key with proper attributes are available. 

The other important part in data publication is providing a pointer to the local (institutional) attribute authority through which the user can ask for a decryption key. In NDN, this is also accomplished by using a namespace. When data is published, the name of the decryption key service (local ledger) is also associated with the data.
\name{/genomics/data/sra1/annotations: encrypted\_by=/genomics/pub\_key/timestamp=<time>/<local\_ledger=/tntech/ledger>}

\subsection{Data Retrieval}
In NDN, data can be retrieved by sending an Interest into the network. For accessing \name{/Genome1/SRA/9605/9609/}, a user simply sends the Interest by that name and receives the encrypted data. The data can come from anywhere, from the publisher, an intermediary repo, or an in-network cache.

\subsection{Decryption Key Generation and Retrieval}
Once the user (let's call her Alice) receives the data, it looks at the annotations in the name. Note that Alice can read the name of the data she received but can not decrypt the payload yet. Alice (or the application Alice is using) needs to request a decryption key ($d_k$). From the forwarding hint in the name, Alice knows she needs to request the $d_k$ from \name{/tntech/ledger}. She sends an Interest to the local ledger in the form of  \name{/tntech/ledger/decryption-key/data:/genomics/data/sra1/<attributes>/Alice-pub-key:/tntech/alice/pub\_key}, where attributes are "PI and PI's graduate students". She also signs the request with her public key, this way the ledger knows the request to be authentic. On receiving this request, TN Tech's ledger looks up Alice's attributes. If Alice is a graduate student working under PI Tom, she will have both attributes in the ledger. The the ledger will sign and forward this request to the NOC. Such a request would look like: \name{/tntech/ledger/decryption-key/data:/genomics/data/sra1/<attributes>/Alice-pub-key:/tntech/alice/pub\_key}, where attributes for Alice are "PI=Tom and Alice is PI's graduate students".

Note that the local ledger can also add additional attributes such as validity period of the requests. On receiving the key, the NOC will generate a decryption key for Alice using the attributes and the ABE keys.

$master_{key}$ + $public_{key}$ + attributes = $decryption-key_{alice}$

The NOC and the local ledger establishes the trust beforehand, and only signed request from the local ledger will create a decryption key. If Alice directly requests the NOC for the decryption key, the NOC will not respond since it does not trust Alice directly.

On receiving the request from the local ledger, the NOC generates and encrypts the decryption key using Alice's public key located at \name{/tntech/alice/pub\_key}. The NOC has two choices to return the key to Alice. The first way is to reply to the local ledger which then returns the key to Alice. The second way is it publishes the key into the NDN network \name{/tntech/ledger/decryption-key\_{alice}/genomics/data/sra1}. Alice then requests the key form the network. Either way, Alice is able to receive the key from the network or the local ledger and decrypt the content. In our implementation, we use the local ledger for distribution.

Note that the key generation and retrieval is a lightweight operation. The application simply stores the decryption key locally and utilizes it in the future. When a new key is needed, the application retrieves a new key. The granularity and lifetime of these keys are established by the NOC, data publishers, and accessing institutes. For example, setting access control attributes over a broader namespace (e.g., /genomics) would require less decryption key generation than setting access control over more specific namesspaces (e.g., /genomics/data/sra).

\subsection{Timing attribute and partial reencryption for Revoking access}

One of the challenging parts of attribute-based encryption is access revocation. Since genomics data is long-lived, utilizing different keys as data is generated is not feasible. On the other hand, re-encrypting data frequently to revoke access is also not cost-effective. There are two distinct threat models that we aim to address. First, a superuser or an intermediary should not be able to access the data even though they can access raw files. Second, a graduate student or other collaborator working on sensitive data should no longer have access to the data after leaving the institution. The problem of key management arises when a user access needs to be revoked. AABAC uses a time based attribute between the local ledger and the NOC to enforce this. 

Here is an example, if a student named Alice requests a key at Time $T_{10}$, the attributes that the local ledger will send to the NOC is "PI and PI's graduate student and Timestamp: $T_{10}$". Note that in NDN, a file is made of a number of smaller Data Packets. If the Data packets was encrypted and published at $T_{9}$, Alice will be able to decrypt the individual packets and reassemble the file. However, if the Data packets of a file is published at $T_{11}$, Alice will not be able to decrypt the data packets. We worked with the genomics scientists to understand the parts of the files that are more critical. Rather than reencrypting the whole file, we periodically reencrypt the file metadata as well as random Data packets and update the data repository. If a file is divided into two Data packets (an example, a file would likely be divided into thousands of Data packets) with timestamps $T_{10}$ and $T_{11}$, and Alice requested a Key at $T_{10}$, Alice can decrypt the packet with timestamp $T_{10}$ but not $T_{11}$. Since the data packets already has a key locator built in, Alice will then request the new key $T_{11}$ to be able to decrypt the Data. Note that Alice only need to request $one$ key with our scheme, a key with $T_{11}$ will be able to decrypt both $T_{10}$ and $T_{11}$. If Alice is no longer authorized to decrypt the data, the local ledger will not forward the request to the NOC to get newer keys to continue decrypting future versions.

The other thing to note here is NDN allows us to set content lifetime on Data packets. Even though NDN caches content in the network, by setting content lifetime to a value lower than reencryption time, we can ensure data with older timestamps will not be available from in-network caches. Our experience shows the encrypting the metadata and a random portion of the data is sufficient to preserve the privacy of data. Even if we perform full reencryption, the average file reencryption requires only around 15 seconds.

\begin{figure}[!ht]
    \centering
     \includegraphics[width=0.47\textwidth]{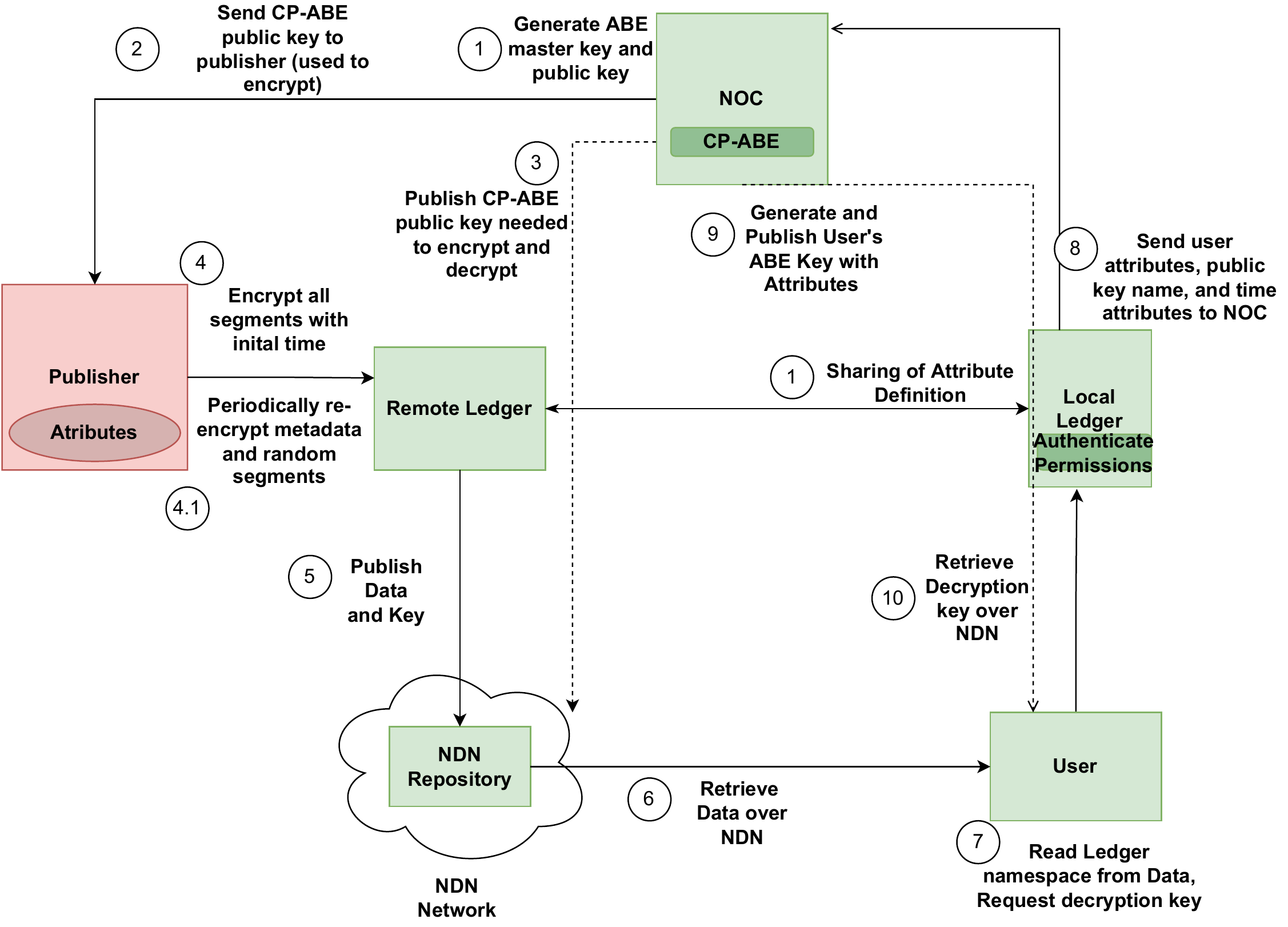}
    \caption{Encryption and Decryption Model}
     \label{fig:encryption_decyption_high_model}
     \vspace{-1em}
\end{figure}

\subsection{System Model}
Figure \ref{fig:encryption_decyption_high_model} provides an overview of the proposed encryption and decryption model outlined in this paper. There are a few main parts: the NOC, the publisher, the user, the remote ledger, the local ledger, and the NDN network that facilities the NDN repository. The first step is understanding the NOC, which generates the CP-ABE keys and maintains control of the master key, which is needed to generate user keys. For a publisher to encrypt and publish data, the first step is to reach out to the NOC and get the CP-ABE public key needed to encrypt the data. The publisher can encrypt data with various attributes in a security policy using this key. For the model to handle the security needed for a changing environment, the publisher will perform multiple encryption solutions. For initial deployment, all data will be encrypted, but in the future, segments of the data will be periodically encrypted again to maintain security and republish to NDN. The next important entity is the user; when the user wishes to decrypt some data from the network, the user contacts the local ledger that authenticates permission and then passes on the request to the NOC to generate a user key with the next time attribute. Once the NOC confirms the request is valid, the NOC will generate a user key with the requested attributes and pass the new user key back to the user through the local ledger. Once the user gets the CP-ABE key containing their attributes, the user can decrypt the data using the key during the accepted time period.

\subsection{A Possible Real-World Scenario}

Having explored the motivation and design for the proposed system in this paper, we aim to demonstrate the system's effectiveness with real-world situations. The distinct scenarios will be primarily based on a hypothetical access framework. The PI is a faculty working for the Biology department. Being a faculty, the PI also has a graduate assistant assisting with project Genome1. The data is also encrypted with periodically changing encryption with increasing sequences to revoke previous access. 
There may be two departments from two different institutions working collaboratively on a single project. The project may include two PIs and multiple graduate assistants working in collaborations. Based on this scenario, the attribute for providing access to the project-related resources would potentially rely on the university's name, the department, and the project. The graduate assistants must be employed under the PIs and assigned to the specific project to access the resources. Anyone outside this group will not be able to update or view the data. The scheme for this would be as follows:

    \subsubsection{Attributes}
    Project, Principal Investigator, University, Department, Role, Time Sequence
    
    \subsubsection{Requested Content Name}
    \name{Genome1/SRA/9605/9609/RNA-Seq/1}

    \subsubsection{User receives content and requests decryption Key Name from Local Ledger}
    \name{/Genome1/SRA/9605/9609/RNA-Seq1/DecryptionKey/Attributes/Name=John Smith/Project=Genome1/University=MIT/Department=Biology/Role=Graduate Assistant/timestamp=1645780366}

    \subsubsection{Access Control Rule and Trust Schema}
    Return user decryption key - [(Project = Genome1) and (((PI = John Smith) and (University = MIT) and (Department = Biology or Department = Computer Science) and (Role = Graduate Assistant)) or ((PI = Jack Robinson) and (University = UCLA) and (Department = Biology or Department = Computer Science) and (Role = Graduate Assistant))) and (timestamp = 1645780366)]
    
    \subsubsection{Example}
    \begin{itemize}
        \item Student 1 - [Project = Genome1; PI = John Smith; University = MIT; Department = Biology; Role = Graduate Assistant; timestamp = 1645780366] - Receives decrypted data
        \item Student 2 - [Project = Genome1; PI = John Smith; University = UCLA; Department = Biology; Role = Graduate Assistant; timestamp = 1645780366] - Does not receive decrypted data
    \end{itemize}

\section{Evaluation}
This section evaluates our framework in terms of performance and overhead. One of the criticisms of attribute-based encryption has been that they are slow. However, genomics data is long-lived, and we show that cost of encryption is manageable. We also show that the per-packet encryption time is low. Since the metadata for SRA genome files is usually small at under 17KB, equivalent to two NDN packets, encrypting the metadata every time is a small cost to keep the data secure. Even when the full file is encrypted, it takes less than 15 seconds to encrypt an average-sized genomics file. We also show that the storage overhead goes up very slightly with the number of attributes, but they do not affect the system performance negatively. 

\subsection{Encryption time with CP-ABE}

\begin{figure}[!ht]
    \centering
     \includegraphics[width=0.4\textwidth]{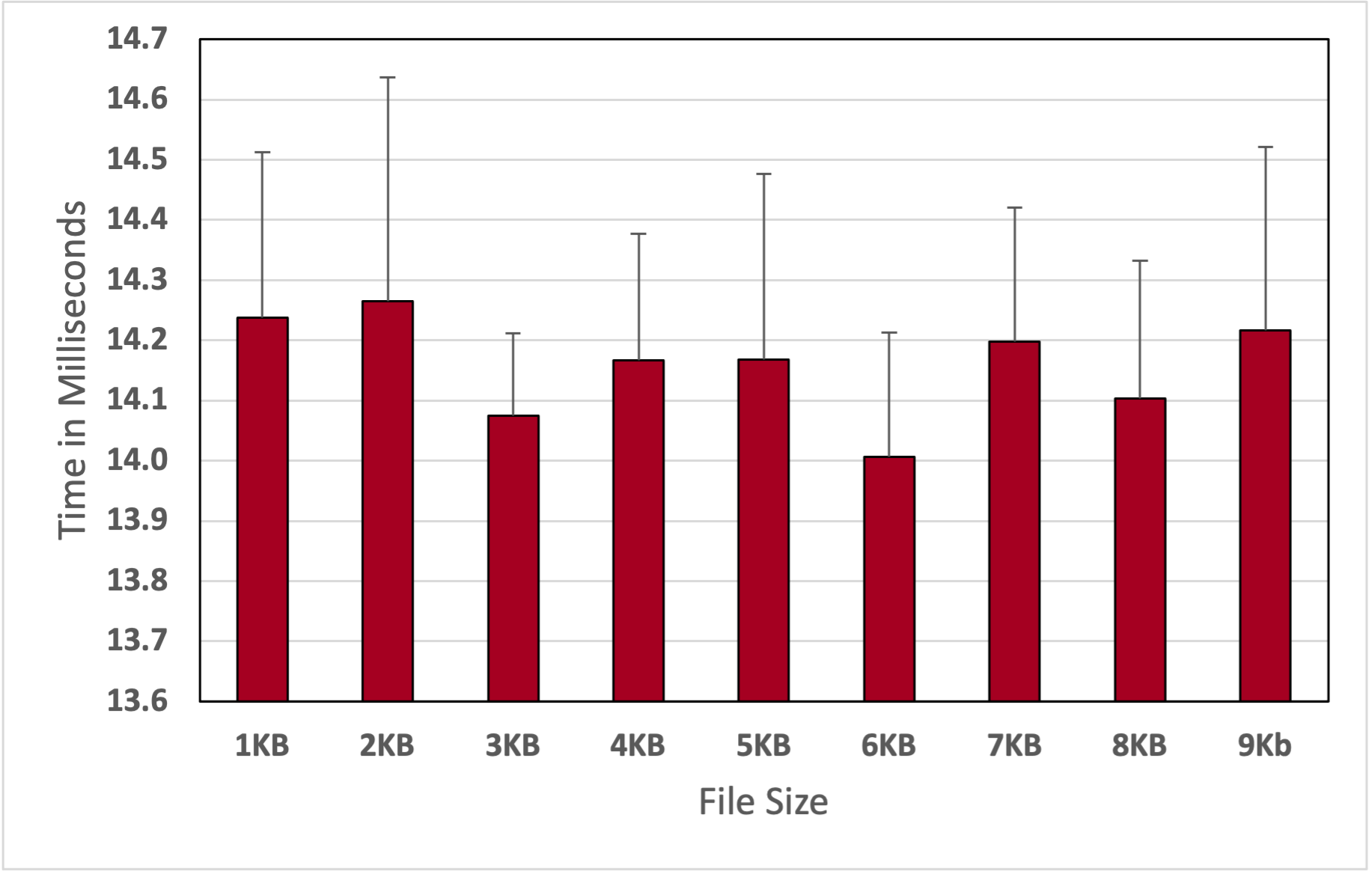}
    \caption{Average Time to Encrypt per Packet Size}
     \label{fig:packet-times}
     \vspace{-1em}
\end{figure}

\begin{figure}
    \centering
     \includegraphics[width=0.4\textwidth]{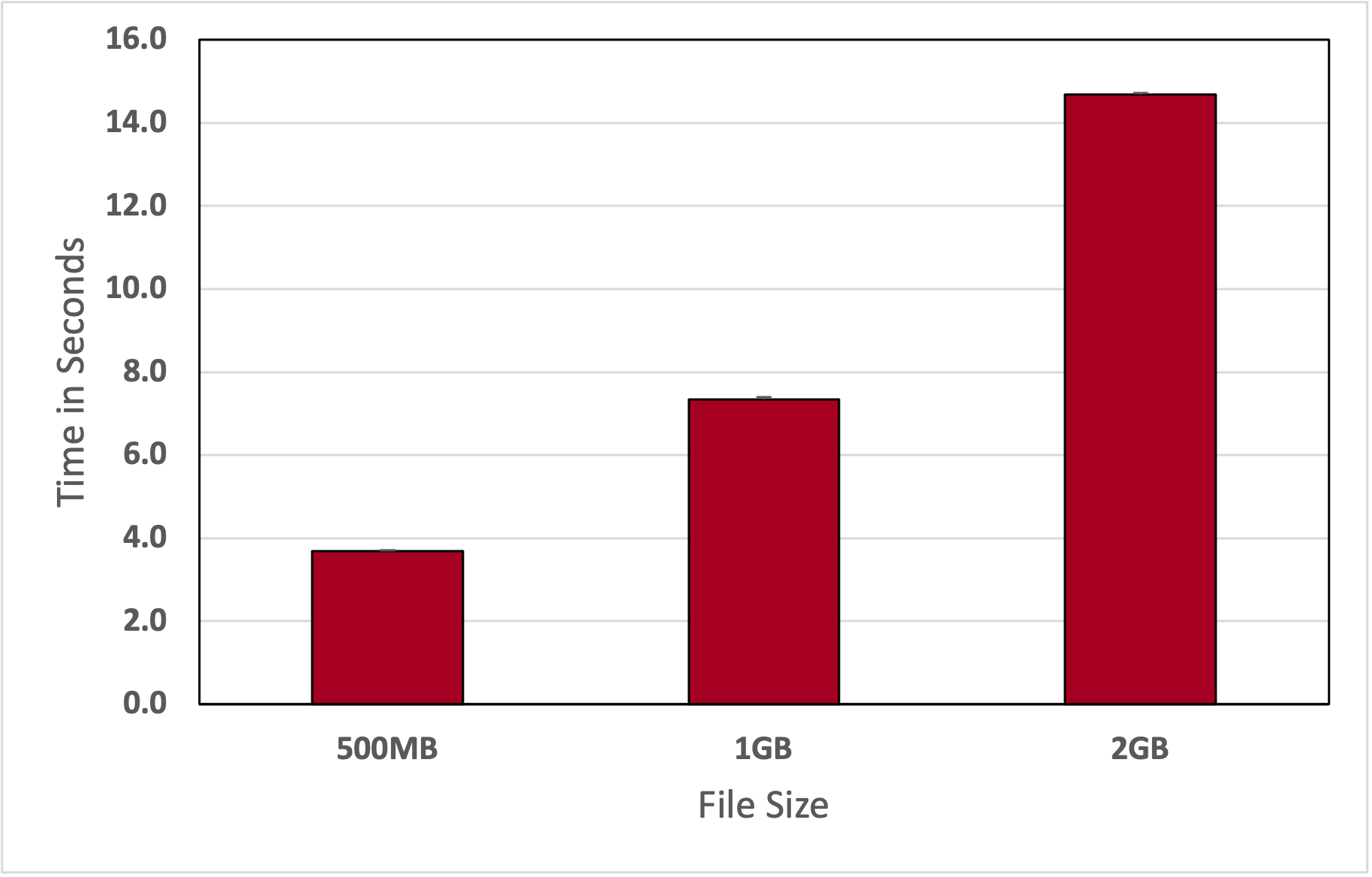}
    \caption{Average Time to Encrypt per File Size}
    \label{fig:large-times}
    \vspace{-2em}
\end{figure}

When working with large data sets that need to be published with a comparatively inefficient encryption algorithm compared to symmetric key encryption, performance is important. Multiple experiments were run to demonstrate that using CP-ABE directly instead of using an intermediate symmetric key encryption is viable. The first test shown in Figure \ref{fig:packet-times} demonstrates the encryption time in milliseconds when working with standard NDN packet sizes that can vary between 0 and 8800 bytes. The figure indicates for these sizes that encryption can be accomplished in between 14 and 15 milliseconds on average over ten runs for each file size. The second experiment shows the total time needed to encrypt the most common genomics samples that average less than 2 GB each. The results for this are shown in Figure \ref{fig:large-times}. The figure shows the average time in seconds for encryption of 500MB, 1GB, and 2GB files when run ten times each.

\subsection{Number of attributes vs encryption time}
\begin{figure}
    \centering
     \includegraphics[width=0.4\textwidth]{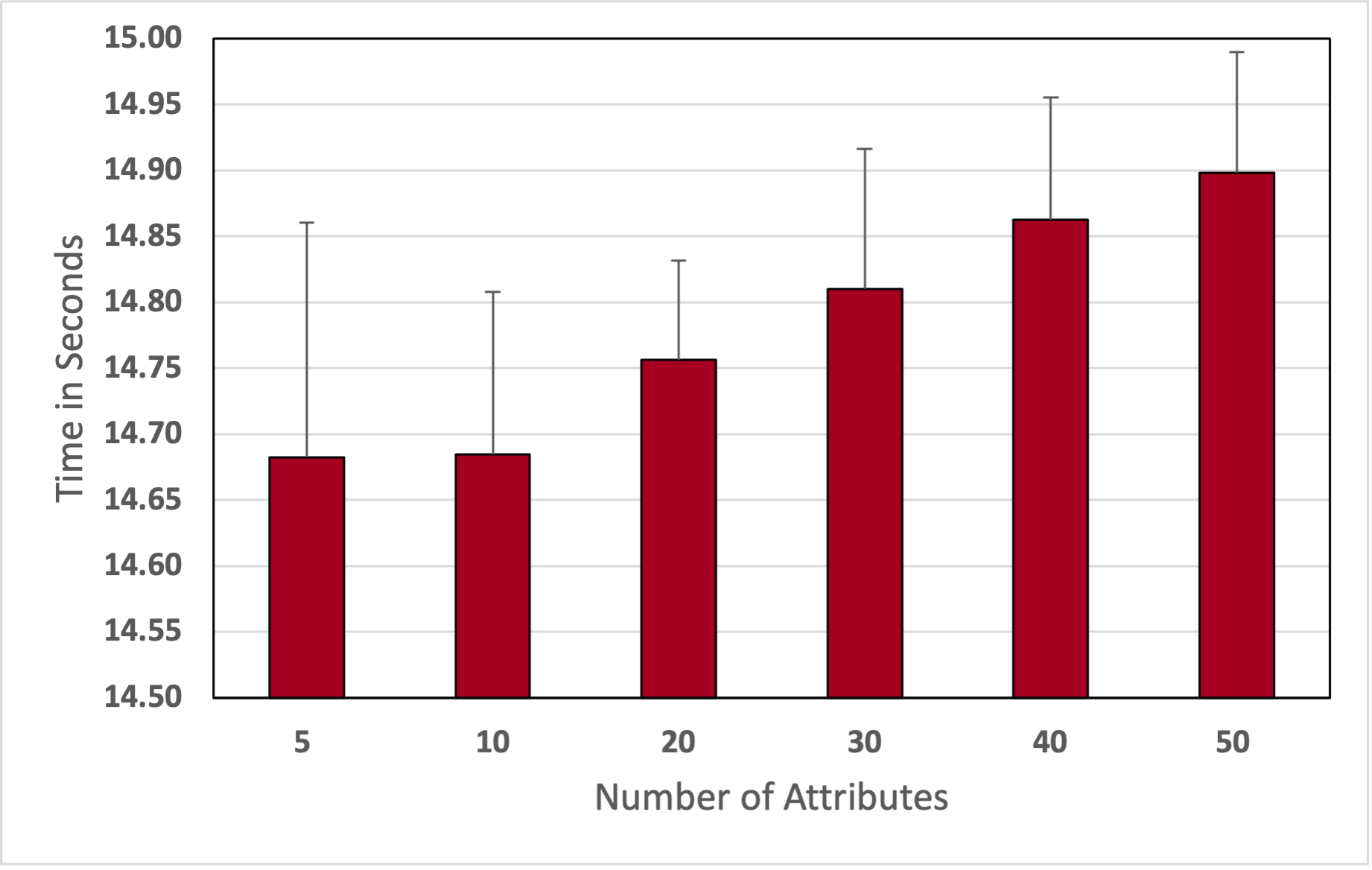}
    \caption{Average Time to Encrypt per Number of Attributes}
     \label{fig:attribute-time}
     \vspace{-1em}
\end{figure}

When working with a scalable encryption algorithm like CP-ABE, concern if added complexity would affect performance was a concern. Experiments were run with a varying number of attributes from five to fifty to determine if this would prove to be a potential problem for some deployments. For this test, the file size was kept constant with an original file equaling 2GB, and the experiment ran ten times for each number of attributes. The results were then plotted, exhibiting the time in seconds with standard deviation for the different number of attributes. As Figure \ref{fig:attribute-time} shows, while increasing the number of attributes does increase the encryption time in a predictable pattern, for the test file, all results average between 14.5 and 15 seconds. 

\subsection{Overhead for NDN}
When working with CP-ABE to secure genomics data, the final concern studied is the file overhead when encrypting. All encryption algorithms will add some overhead, but the degree of overhead can vary between algorithms. Experiments were run to determine the degree of file overhead for a 2GB file depending on the number of attributes to alleviate the concern of the significance of the wasted file overhead. The results show the overhead is very predictable and consistent when using the same size input file, in this case, 2 GB, and only changing the number of attributes. Figure x demonstrates the overhead in bytes over the original 2GB file for the number of attributes varying from five to fifty over ten runs each. While this shows that increasing the number of attributes will increase the file overhead, this overhead will require only one or two extra NDN packets for delivery when working with less than fifty attributes.

\begin{figure}
    \centering
     \includegraphics[width=0.4\textwidth]{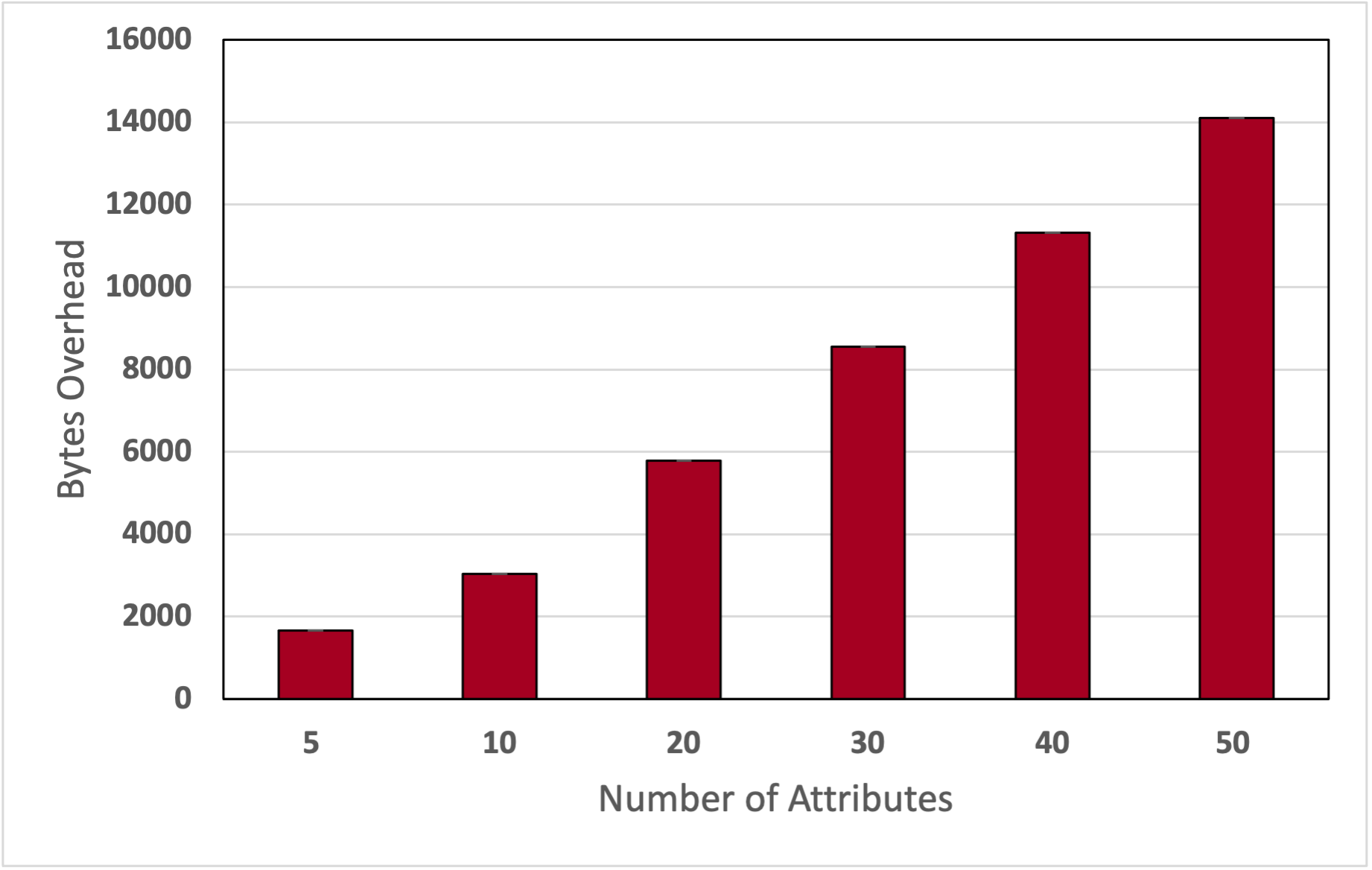}
    \caption{Average File Overhead Size for 2GB for Number of Attributes}
     \label{fig:large-overhead}
     \vspace{-1em}
\end{figure}


\section{Conclusion and Future Work}
The vast availability of genomics data has highlighted the need to ensure security and privacy when sharing healthcare information. Access control mechanisms based on roles and attributes are key factors that must be taken under consideration to facilitate such assurances. The goal of this paper is to introduce attribute-based access control for genomics data. 
We plan to implement our prototype and integrate our work with an actual genomic workflow and evaluate its performance in the near future.

\bibliographystyle{abbrv}
\bibliography{refs}

\end{document}